\documentclass{article}

\usepackage[utf8]{inputenc}
\usepackage[T1]{fontenc}
\usepackage{graphicx}
\usepackage{amsmath, amssymb}
\usepackage{natbib}
\usepackage{geometry}
\usepackage{lipsum}
\usepackage{booktabs}
\usepackage{threeparttable}
\usepackage{float}
\usepackage{placeins}
\usepackage{caption}
\usepackage{siunitx, tabularx, makecell}
\usepackage{ragged2e}
\begin{document}
\justifying

\begin{titlepage}
    \centering
    \vspace*{4cm}

    {\Huge\bfseries
    Rethinking Portfolio Risk: Forecasting Volatility Through Cointegrated Asset Dynamics \par}

    \vspace{2cm}

    {\Large Gabriele Casto\par}

    \vspace{1cm}

    {\large September 2025\par}

    \vspace*{\fill}
\end{titlepage}

\section{Introduction}

Volatility plays a central role in modern portfolio theory, as it is the dominant measure for quantifying financial risk. Since Markowitz's initial work in 1952 \citep{Markowitz1952}, portfolio optimization has been based on the assumption that asset returns belong to the family of elliptical distributions, so that the first two moments could provide a reliable approximation for risk and return. Although this approach is theoretically elegant, it faces significant challenges in terms of practical applicability: when dealing with high-dimensional portfolios or in periods of financial turmoil, the covariance matrix tends to become unstable, poorly conditioned, or difficult to estimate precisely.

\hspace{0.5em}

To address such limitations, recent studies have proposed new frameworks for estimating volatility, particularly those founded upon the time-series nature of individual securities in comparison with benchmark indices \citep{Bollerslev1986,Engle2002}. In this work, we suggest and examine two new ratio-based indicators of risk: the Historical Volatility Ratio (HVR) and the Dynamic Volatility Ratio (DVR). They both compare the relative risk of a security with a benchmark index but differ in nature. The HVR relies on historical realized volatility \citep{Andersen2001}, whereas the DVR employs the conditional variance predictions of the GARCH-style models in an attempt to construct a forward-looking view of relative risk \citep{Bollerslev1986,Engle2002}.

\hspace{0.5em}

We argue here that these ratios should not merely be seen as statistical tools but as relative risk intensity measures rooted in theory and possessing desirable properties such as scale invariance, interpretability, and stationarity in many cases. Notably, we also show here that, in a wide range of horizons, the stock volatilities are cointegrated with the corresponding index volatilities, so we can envisage these ratios as mean-reverting stochastic processes. This opens the possibility of addressing portfolio risk estimation without the need for a direct computation of the classical covariance matrix.

\hspace{0.5em}

The paper contributes to the literature in the following ways:
\begin{itemize}
    \item First, it provides a conceptual definition of HVR and DVR and the theoretical implications of these risk measures.
    \item Second, it examines whether there is statistical evidence of stationarity for these ratios and cointegration among volatility series \citep{EngleGranger1987,Johansen1991}.
    \item Third, we formulate a theoretical framework in terms of which portfolio volatility can be expressed and forecast by a Vector Error Correction Model (VECM) given the assumption of cointegration among volatility series \citep{Johansen1991}.
    \item Finally, we substantiate our methodology in an empirical section (Section 7) where we validate the ratio-based approach and demonstrate the superiority of the VECM over the common covariance-based volatility estimates with S\&P 500 constituents.
    \item In the near future, I expect to expand this initial study with new frameworks to identify stable asset volatility structures across different markets and horizons and to study the connection between these structures and market efficiency.
\end{itemize}

The subsequent sections of this paper are organized as follows. Section 2 will briefly mention the theoretical foundations of classical volatility modeling and its shortcomings (Markowitz covariance matrix) and define volatility ratios. Section 3 will elaborate on the statistical properties of HVR and DVR. Section 4 establishes a Vector Error Correction Model (VECM) framework under the assumption of cointegration of asset volatilities. Section 5 will revisit the Markowitz optimization framework and CAPM in light of ratio-based volatility. Section 6 examines the limitations of the proposed framework. Section 7 provides empirical findings, while Section 8 offers concluding remarks.

\section{Theoretical Foundations of Volatility in Portfolio Construction}

\subsection{Limitations of Covariance-Based Portfolio Risk Estimation}

In the classical Markowitz model, portfolio risk is defined as the variance of the portfolio returns and is estimated via the sample covariance matrix. Formally, given a portfolio of $n$ assets with return vector $\mathbf{r}_t$ and weight vector $\mathbf{w}$, the variance of the portfolio is given by:
\[
\sigma_p^2 = \mathbf{w}^\top \boldsymbol{\Sigma} \mathbf{w}
\]
where $\boldsymbol{\Sigma}$ is the covariance matrix $n \times n$ of the returns of the assets. The classical approach as defined by Markowitz suffers from certain limitations.

\begin{itemize}
    \item \textbf{High dimensionality:} Constructing covariance matrices becomes statistically and computationally expensive as $n$ grows \citep{LedoitWolf2003}.
    \item \textbf{Instability and noise:} Realized covariance matrices tend to be unstable and sensitive to market regimes when tested out-of-sample \citep{Michaud1989}.
    \item \textbf{Lack of interpretability:} Covariance matrices lack intuitive economic interpretation, especially in cross-sectional or regime-switching environments.
    \item \textbf{Static nature of forecast:} The Markowitz model provides a static risk forecast that does not properly incorporate dynamic relations between input variables; this approach forces a static and single-horizon model into a dynamic volatility environment.
\end{itemize}

These shortfalls have led researchers to explore alternative risk estimators, focusing on univariate models for volatility that are easier to interpret while showing greater stability over time.

\subsection{Ratio-Based Volatility Modeling: A New Perspective}

This paper proposes an alternative approach to estimate asset risk, based on its relative volatility to a market index. We construct a volatility ratio, computed with historical (realized) volatility or conditional (predicted) volatility, that can then be used to express asset risk in relative rather than absolute terms.

\hspace{0.5em}

Let $\sigma_{i,t}^{(k)}$ denote the average volatility of asset $i$ over $k$ units of time (days, minutes, etc.), computed at time $t$, and $\sigma_{m,t}^{(k)}$ the average volatility of the corresponding benchmark index on the same time interval $k$ and moment of estimation $t$. We define:

\begin{itemize}
    \item \textbf{Historical Volatility Ratio (HVR):}
    \[
    \text{HVR}_{i,t}^{(k)} = \frac{\hat{\sigma}_{i,t}^{(k)}}{\hat{\sigma}_{m,t}^{(k)}}
    \]
    where $\hat{\sigma}_{i,t}^{(k)}$ is a realized standard deviation of the asset returns over the past $k$ days.
    
    \item \textbf{Dynamic Volatility Ratio (DVR):}
    \[
    \text{DVR}_{i,t}^{(k)} = \frac{\sigma_{i,t+1|t}^{(k)}}{\sigma_{m,t+1|t}^{(k)}}
    \]
    where $\sigma_{i,t+1|t}^{(k)}$ is the one-step-ahead conditional standard deviation forecast from a GARCH-type model.
\end{itemize}

These ratios offer several advantages:

\begin{itemize}
    \item They are scale-free; therefore, they are comparable across time and different assets.
    \item They capture the relative risk intensity of an asset compared to a benchmark, providing information about the sensitivity of the asset returns to broader market conditions.
    \item Empirical analysis presented in this paper (Section 7) shows that assets' volatilities are often stationary and in many cases cointegrated with market volatility, which supports their use in econometric modeling.
\end{itemize}

This approach shifts the focus of risk modeling from absolute variance structure to relative volatility dynamics, which are often more stable and interpretable in practice. This motivates the integration of these ratios into a cointegration-based volatility forecasting framework.

\section{Statistical Properties of Volatility Ratios}

\subsection{Stationarity of Relative Volatility Measures (ADF)}

In time series econometrics, stationarity is a desirable feature of any variable used in prediction or modeling. In a stationary series, the mean, variance, and autocovariance structure are time-invariant, which ensures stable statistical inference.

Let $\text{HVR}_{i,t}^{(k)}$ and $\text{DVR}_{i,t}^{(k)}$ denote the Historical and Dynamic Volatility Ratios, respectively. We are interested in testing whether:

\[
\text{HVR}_{i,t}^{(k)} \sim I(0), \quad \text{and/or} \quad \text{DVR}_{i,t}^{(k)} \sim I(0)
\]

\hspace{0.5em}

Preliminary empirical results (Section 7) suggest that, for most stocks in the S\&P 500 index, such ratios exhibit stationary behavior across a variety of daily and intraday rolling windows ($5, 10, 30, 90$ days and $5, 10, 30, 90$ minutes).

While individual asset volatilities display trends and breaks in structure, relative volatility within a market tends to revert to a stable level in the long run. This tendency reveals the presence of enduring risk premia or beta-like patterns.

\subsection{Cointegration with Benchmark Volatility}

Besides stationarity, another essential feature that supports the use of the defined volatility ratios is the presence of cointegration between volatilities of asset log-returns and those of their respective index log-returns. We define the volatility of the log-returns of an asset $i$ and the benchmark $m$ as $\sigma_{i,t}^{(k)}$ and $\sigma_{m,t}^{(k)}$. We are interested in testing the stationarity of the residuals $\varepsilon_t$ of the equation:

\hspace{0.5em}

\[
\sigma^{(k)}_{i,t} = \alpha + \beta \cdot \sigma^{(k)}_{m,t} + \varepsilon_t
\]

Formally:

\[
\varepsilon_t \sim I(0)
\]

\hspace{0.5em}

This evidence would imply that the volatility of the asset and that of the respective benchmark share a common stochastic trend, which validates the construction of a ratio-based measure such as HVR and DVR.

\hspace{2em}

The presence of cointegration, as shown in Section 7, offers justification for volatility ratios being bounded and possessing mean-reverting properties, therefore allowing Vector Error Correction Models (VECM) to be used in volatility modeling as well as in portfolio forecasting. Such an approach allows us to separate the short-term dynamics, obtained by differencing, and the behavior in the long-term equilibrium, encoded in the cointegrating vector. Under these assumptions, this model is capable of providing a more robust and forward-looking structure over standard variance estimation.

\newpage

\section{A VECM-Based Forecasting Framework for Portfolio Volatility}

\subsection{Motivation for a Cointegration-Based Approach}

The classical Markowitz model relies on the covariance matrix of returns (or log-returns) of assets to estimate portfolio risk. However, as previously mentioned, estimating high-dimensional covariance matrices poses a serious challenge for the reliability of forecasts as they tend to be unstable and noisy, especially in periods of turmoil \citep{LedoitWolf2004}.

\hspace{0.5em}

Instead of directly modeling raw variances and covariances, we recommend simultaneous modeling of log-volatilities of all portfolio assets via a multivariate cointegration approach. In particular, we assume that log-volatility series of individual stocks are collectively cointegrated, reflecting the presence of long-run equilibrium relationships among them. This is supported by empirical results provided in Section 7, as:

\begin{itemize}
\item For each asset $i_n$, if its volatility $\sigma_{i,t}^{(k)}$ is cointegrated with the benchmark index volatility $\sigma_{m,t}^{(k)}$, then the set of asset volatilities shares a common stochastic trend with the benchmark, implying cointegration among all volatilities.
\item If two strictly positive time series are cointegrated, it follows that their logarithmic transformation is also cointegrated. This follows from the fact that the logarithmic function is monotonic and differentiable, and thus preserves the order and the linear relationships required for cointegration.
\end{itemize}

These considerations allow us to employ a Vector Error Correction Model (VECM) to forecast the $(t+h)$ future values of volatilities for a set of $n$ assets in a unified system. Using the Vector Error Correction Model (VECM), we can decompose the dynamics of each volatility series into:

\begin{itemize}
    \item Short-run fluctuations (via first differences),
    \item Long-run adjustment toward equilibrium (via the error correction term).
\end{itemize}

This allows for more robust and interpretable volatility forecasts than traditional approaches.

\subsection{VECM Specification}
Let 
$\mathbf{h}_t = \left[ \log(\sigma_{1,t}), \log(\sigma_{2,t}), \dots, \log(\sigma_{n,t}) \right]^\top$
 be an $n$-dimensional vector of log-volatilities for $n$ assets. We assume that there exist $r < n$ linearly independent cointegrating vectors such that $\beta^\top \mathbf{h}_t \sim I(0)$ \citep{Johansen1991}.
 Assuming that the $n$ log-volatilities are cointegrated and share a long-run equilibrium structure, we impose a cointegration rank of $r = n-1$.
 Then the VECM representation is:

\[
\Delta \mathbf{h}_t = \Pi \mathbf{h}_{t-1} + \sum_{j=1}^{p-1} \Gamma_j \Delta \mathbf{h}_{t-j} + \varepsilon_t
\]

where:

\begin{itemize}
    \item $\Pi = \alpha \beta^\top$ encodes the long-run relationships (error correction),
    \item $\Gamma_j$ captures the short-run dynamics,
    \item $\varepsilon_t \sim \mathcal{N}(0, \Sigma)$ are innovations.
\end{itemize}

Note that, in the case of two assets, the cointegrating relationship can be written as:
\[
\beta^\top \mathbf{h}_t = \log(\sigma_{1,t}) - \log(\sigma_{2,t})
\]

This corresponds to the logarithm of the Historical Volatility Ratio between the two assets.

\subsection{Forecasting Portfolio Volatility with VECM}

We can implement this model to forecast the portfolio volatility as follows:

\begin{enumerate}
    \item Estimate the VECM on the log-volatilities of all $n$ assets in the portfolio, assuming a cointegration rank of $r = n-1$.
    
    \item Forecast future asset volatilities $\hat{\sigma}_{i,t+1}, \hat{\sigma}_{i,t+2}, \dots, \hat{\sigma}_{i,t+h}$ over horizon $h$ by back-transforming the VECM forecast of $\log(\sigma_{i,t+h})$.

    \item For each forecast horizon $j = 1, \dots, h$, construct a diagonal matrix $\hat{D}_{t+j}$ with the asset volatilities, and compute the portfolio variance as:
    \[
    \hat{\sigma}_{\text{portfolio}, t+j}^2 = \mathbf{w}^\top \hat{D}_{t+j} \cdot \hat{\rho}_{t} \cdot \hat{D}_{t+j} \mathbf{w}
    \]

    \item Average the forecasted daily portfolio variances over the $h$-day window and compute the expected portfolio volatility as:
    \vspace{1em}
    \[
    \hat{\sigma}_{\text{portfolio}, t}^{(h)} = \sqrt{ \frac{1}{h} \sum_{j=1}^{h} \hat{\sigma}_{\text{portfolio}, t+j}^2 }
    \]
\end{enumerate}

\subsection{Advantages Over Classical Approaches}

Compared to the sample covariance method, the VECM-based model provides:
\begin{itemize}
    \item \textbf{Stability:} Cointegration reduces noise and isolates persistent relationships.
    \vspace{1em}
    \item \textbf{Forecastability:} The VECM is forward-looking by construction, decomposing volatility dynamics into short-run fluctuations and long-run adjustments. This allows us to overcome the static nature of the covariance matrix and to incorporate dynamic relations between variables.
\end{itemize}

\hspace{0.25em}

\section{Theoretical Implications in Classical Portfolio Models}

\subsection{Volatility Ratios in Covariance Matrix Estimation}

Under the assumption of cointegration, we expect the ratios $\text{HVR}_{i,t}$ and $\text{DVR}_{i,t}$ to be stationary. This implies that changes in $\sigma_{i,t}$ are proportionally mirrored in $\sigma_{m,t}$. Therefore, we can approximate future variances as:
\[
\sigma^2_{i,t} \approx (\text{HVR}_{i,t})^2 \cdot \sigma^2_{m,t}
\]

\hspace{0.5em}

Or, alternatively, using the Dynamic Volatility Ratio (DVR) as:

\hspace{0.5em}

\[
{\sigma_{i,t+1|t}^{2}} \approx \text{DVR}_{i,t}^{2} \cdot {\sigma_{m,t+1|t}^{2}}
\]

\hspace{0.5em}

Therefore, the covariance between two assets $i$ and $j$ can be expressed as:

\[
\text{Cov}(i,j)_{t} \approx \rho_{i,j} \cdot \text{HVR}_{i,t} \cdot \text{HVR}_{j,t} \cdot \sigma^2_{m,t}
\quad \text{,} \quad
\text{Cov}(i,j)_{t+1} \approx \rho_{i,j} \cdot \text{DVR}_{i,t} \cdot \text{DVR}_{j,t} \cdot {\sigma_{m,t+1|t}^{2}}
\]

where $\rho_{i,j}$ represents the correlation coefficient between the log-returns of asset $i$ and asset $j$. This formulation leads to a covariance matrix $\Sigma_t$ of the form:

\[
\Sigma_t \approx \sigma^2_{m,t} \cdot \big(\text{HVR}_t\, \text{HVR}_t^\top\big) \circ R
\quad \text{,} \quad
\Sigma_{t+1} \approx \sigma_{m,t+1|t}^2 \cdot \big(\text{DVR}_{t+1|t}\, \text{DVR}_{t+1|t}^\top\big) \circ R
\]

where $R$ is the correlation matrix of asset returns and $\circ$ denotes the Hadamard (element-wise) product.

Although mathematically equivalent to the classical covariance computation, the decomposition of the covariance matrix with HVR offers a different perspective on estimating the covariance matrix as it separates market volatility, relative asset exposures to market fluctuations (HVR) and the correlation matrix.

At the same time, computing the covariance matrix by using the DVR allows for better estimation in periods of market turmoil if the assumptions of the model are verified, as the GARCH model can provide a better approximation of the market volatility and correlation estimates tend to be less noisy than the covariances.
When computing the DVR, we should employ ARIMA models to avoid the direct estimation of $\sigma_{i,t+1|t}^{2}$.

\subsection{Volatility Ratios in CAPM Estimation}

We can employ the Historical Volatility Ratio (HVR) to provide a different expression of the CAPM model.

\hspace{0.5em}

We can define $\beta_t$ as:
\[
\beta_{i,t} \approx \mathrm{Corr}(i,m) \cdot \text{HVR}_{i,t}
\]

It follows that:
\[
\mathbb{E}[R_i] - R_f = \mathrm{Corr}(i,m) \cdot \text{HVR}_{i,t} \cdot \left(\mathbb{E}[R_m] - R_f\right)
\]

This construction directly links the excess return of asset $i$ to its relative volatility, as captured by the Historical Volatility Ratio, scaled by a correlation factor. Together, these elements reflect the asset's representative responsiveness to changes in the market as well as its sensitivity to systematic risk exposure.

\subsection{Naive Volatility Model}

In Section 7 we will test whether a linear model can be suitable to capture asset volatility based on the volatility of the respective index for the same $n$-day horizon.
The models are constructed as follows:

\[
\sigma_{i,t}^{(k)} = \alpha + \beta \cdot \sigma_{m,t}^{(k)} + \varepsilon_t
\quad \text{,} \quad
\sigma_{i,t}^{(k)} = \beta \cdot \sigma_{m,t}^{(k)} + \varepsilon_t
\]

\section{Limitations and Future Work}

While the volatility ratio-based framework does have its advantages in terms of performance and interpretability, it also comes with certain limitations.

\hspace{0.5em}

First, the cointegration assumption of asset volatilities, as experimentally justified in Section 7 of this work, may not be universally applicable over horizons, industries, or during periods of structural reforms. Disruptions in such an assumption may reduce prediction accuracy and undermine the theoretical robustness of the VECM as a research tool.

\hspace{0.5em}

Also, while our scheme circumvents a direct estimate of the entire covariance matrix, it does entail correlation matrix estimation. Correlation matrices in high-dimensional portfolios could become prone to instability and sensitivity to noise in a way that may undermine reliability in estimating volatility portfolios.

\hspace{0.5em}

Lastly, although the VECM is meant for prediction and encompasses both long-run equilibrium and a short-run departure, its performance relies upon lag selection, sample space, as well as underlying assumptions about distributions.

\hspace{0.5em}

Future work will investigate the strength of the proposed framework for various asset classes, markets, and macroeconomic environments. Additionally, as proven by the empirical results, it is easier and more precise to estimate volatility ratios and equilibria across assets than individual volatilities. These findings will be further investigated to generate a framework that identifies volatility structures across assets and their implications for market efficiency.

\section{Empirical Results}

This section contains an empirical assessment of the volatility ratio-based framework for forecasting portfolio risk. Using a comprehensive dataset of all S\&P 500 constituents at both daily and intraday frequencies, we validate the ability of the Vector Error Correction Model (VECM) to forecast portfolio volatility and compare it with the conventional sample covariance method. This analysis is based on daily data from 2018/09/01 to 2025/09/01 and intraday data (each minute) from 2024/09/01 to 2025/09/01.

\hspace{0.5em}

The empirical part is carried out in multiple steps. First, we construct rolling historical volatilities \citep{Andersen2001} for each asset and the index over horizons of 5, 10, 30, and 90 minutes, and 5, 10, 30, and 90 days, and test whether the Naive Volatility Model (Section 5.3) is effective in capturing the relationship between an asset’s volatility and that of its benchmark index. Then, we generate the rolling HVRs for each asset under the proposed horizons and test whether these ratios show stationarity and whether the volatility of each asset shows statistical evidence of cointegration with the index. We then proceed to study the distribution of the HVRs over the various time horizons and the effectiveness of ARIMA models in forecasting the HVRs. Finally, we estimate a multivariate VECM and generate multi-step-ahead forecasts of log-volatilities. These forecasts are back-transformed to standard deviations, combined with correlation estimates, and used to compute expected portfolio volatility.

\hspace{0.5em}

For benchmarking purposes, we also compute the classical Markowitz portfolio volatility using the sample covariance matrix over the same rolling windows. We compare both methods with the realized volatility of the portfolio over the forward horizon. We use the Mean Absolute Percentage Error (MAPE) as the primary metric, being aware that, with small denominators, the MAPE is likely to show high values.

This empirical implementation allows us to test whether the cointegration-based volatility forecasting framework provides superior accuracy and stability relative to traditional methods, particularly in high-dimensional settings and across different time horizons.

\subsection{Naive Volatility Model}
As mentioned in Section 5.3, we have built a naive model to interpret asset volatility as a function of its relative index volatility. Computing the model at horizons of 5, 10, 30, 90 minutes and 5, 10, 30, 90 days, we observe the following results:

\begin{table}[H]
\centering
\caption{Outputs of Model 1 and Model 2 for the $h$-minute horizon}
\label{tab:vol_summary_min}
\begin{threeparttable}
\begin{tabular}{lcccc}
\toprule
 & 5 min & 10 min & 30 min & 90 min \\
\midrule
\multicolumn{5}{l}{\textit{Model 1: No intercept}} \\
\% $\beta$ significant ($p<0.05$) & 100\% & 100\% & 100\% & 100\% \\
Mean Adj. $R^2$ (\%) & 54.7 & 60.7 & 66.2 & 69.8 \\
Mean MAPE (\%)       & \multicolumn{1}{c}{--} & \multicolumn{1}{c}{--} & 36.7 & 34.7 \\
Average AIC       & -1,138,113.1 & -1,153,403.2 & -1,169,318.0 & -1,180,629.5 \\
Average BIC       & -1,138,094.1 & -1,153,384.3 & -1,169,299.1 & -1,180,610.6 \\
\midrule
\multicolumn{5}{l}{\textit{Model 2: With intercept}} \\
\% $\beta$ significant ($p<0.05$) & 100\% & 100\% & 100\% & 100\% \\
Mean Adj. $R^2$ (\%) & 35.5 & 40.3 & 44.2 & 44.5 \\
Mean MAPE (\%)       & \multicolumn{1}{c}{--} & \multicolumn{1}{c}{--} & 36.9 & 34.1 \\
Average AIC       & -1,155,535.1 & -1,173,009.4 & -1,192,136.1 & -1,209,736.1 \\
Average BIC       & -1,155,506.7 & -1,172,980.9 & -1,192,107.7 & -1,209,707.6 \\
\bottomrule
\end{tabular}
\begin{tablenotes}
\footnotesize
\item Note: Each statistic summarizes regressions of stock volatility on S\&P 500 volatility across all stocks in the sample. MAPE and Adj. $R^2$ are expressed in percentage terms. Data from 2024/09/01 to 2025/09/01
\item Missing MAPE (\%) for the 5- and 10-minute horizons are due to the presence of zeros in the series, which inflates these values toward infinity.
\end{tablenotes}
\end{threeparttable}
\end{table}

\vspace{2em}

\justifying

Overall, we observe the superiority of Model 1 over Model 2. These results justify the utilization of HVRs as a relative risk metric. Furthermore, these models show a tendency to improve the fit and descriptive capabilities as $h$ increases. This behavior is expected as too-short samples tend to be noisy and a naive linear model will perform poorly, while, when tested on longer time horizons, the model shows a better fit as the noise tends to decrease.

\hspace{1em}

Moving to a daily horizon, we observe the following results:

\vspace{2em}

\begin{table}[H]
\centering
\caption{Outputs of Model 1 and Model 2 for the $h$-day horizon}
\label{tab:vol_summary_day}
\begin{threeparttable}
\begin{tabular}{lcccc}
\toprule
 & 5 days & 10 days & 30 days & 90 days \\
\midrule
\multicolumn{5}{l}{\textit{Model 1: No intercept}} \\
\% $\beta$ significant ($p<0.05$) & 100\% & 100\% & 100\% & 100\% \\
Mean Adj. $R^2$ (\%) & 75.1 & 82.1 & 88.5 & 92.9 \\
Mean MAPE (\%)       & 43.9 & 33.8 & 26.3 & 21.6 \\
Average AIC       & -11,051 & -11,637 & -12,343 & -12,808 \\
Average BIC       & -11,040 & -11,626 & -12,332 & -12,798 \\
\midrule
\multicolumn{5}{l}{\textit{Model 2: With intercept}} \\
\% $\beta$ significant ($p<0.05$) & 100\% & 100\% & 100\% & 100\% \\
Mean Adj. $R^2$ (\%) & 43.8 & 52.4 & 63.1 & 71.0 \\
Mean MAPE (\%)       & 48.0 & 32.7 & 22.6 & 16.3 \\
Average AIC       & -11,515 & -12,228 & -13,203 & -13,947 \\
Average BIC       & -11,498 & -12,212 & -13,187 & -13,931 \\
\bottomrule
\end{tabular}
\begin{tablenotes}
\footnotesize
\item Note: Each statistic summarizes regressions of stock volatility on S\&P 500 volatility across all stocks in the sample. MAPE and Adj. $R^2$ are expressed in percentage terms. Data from 2018/09/01 to 2025/09/01
\end{tablenotes}
\end{threeparttable}
\end{table}

\vspace{2em}

\justifying

On a daily basis, we notice mixed results for the fit of these models: the Adj. $R^2$ shows a greater explanatory power for Model 1, but MAPE, AIC and BIC reveal a slightly better fit of Model 2. This result is justified as the volatility of a stock reacts both to changes in market conditions and to a "base-level" volatility that arises due to idiosyncratic factors not captured by changing market conditions.

\vspace{3em}

\subsection{Rolling HVR visualization}
As mentioned above, we compute Historical Volatility Ratios as  $\text{HVR}_{i,t}^{(k)} = {\hat{\sigma}_{i,t}^{(k)}} / {\hat{\sigma}_{m,t}^{(k)}}$. We are interested in studying the properties of this ratio, especially whether we can statistically prove its stationarity.
Before presenting the results obtained, it is helpful to visualize an example of HVR computed on the 5, 10, 30, 90 minutes and 5, 10, 30, 90 days horizons.
Displayed below are the HVR line plots for Apple Inc. across the examined horizons.

\begin{figure}[H]
    \centering
    \begin{tabular}{cc}
        \includegraphics[width=0.45\textwidth]{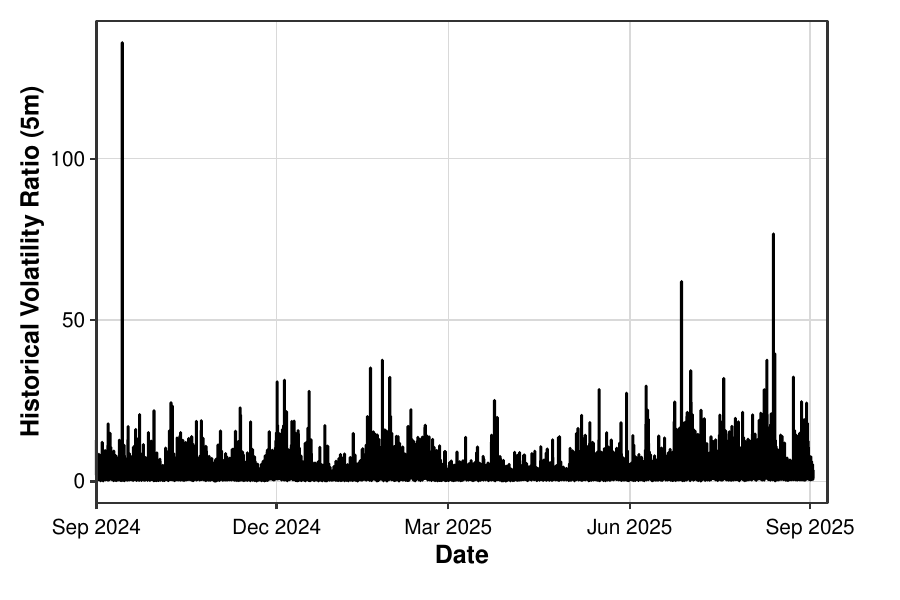}  &
        \includegraphics[width=0.45\textwidth]{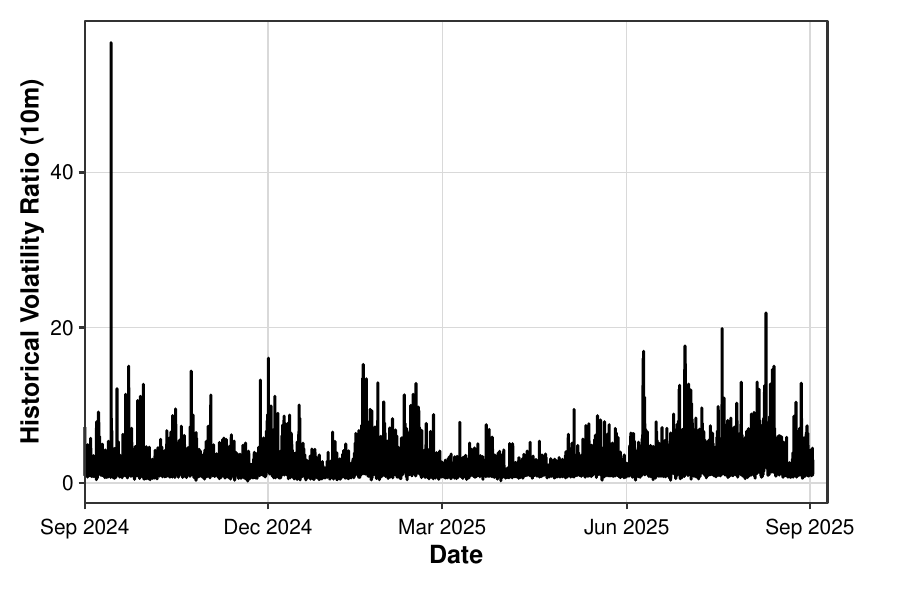} \\
        \includegraphics[width=0.45\textwidth]{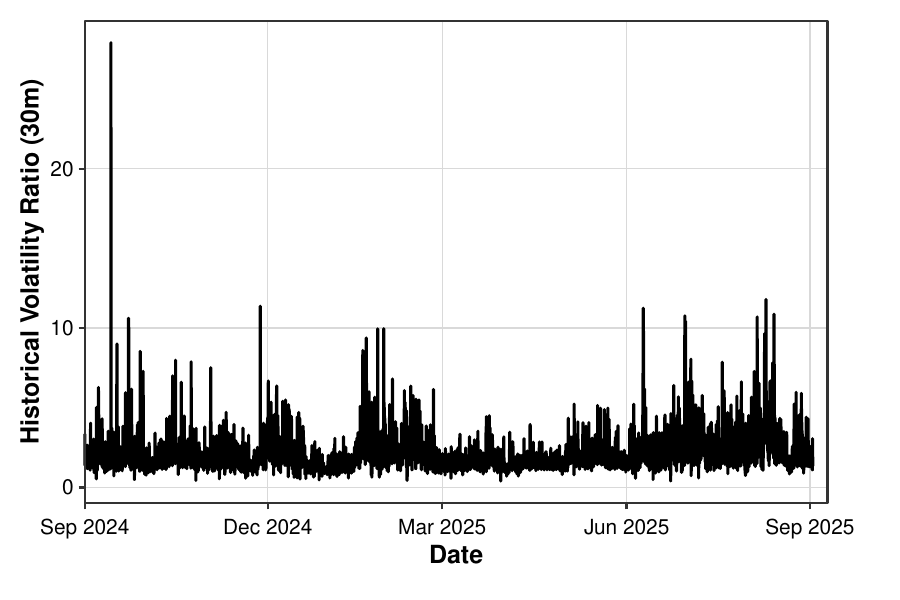} &
        \includegraphics[width=0.45\textwidth]{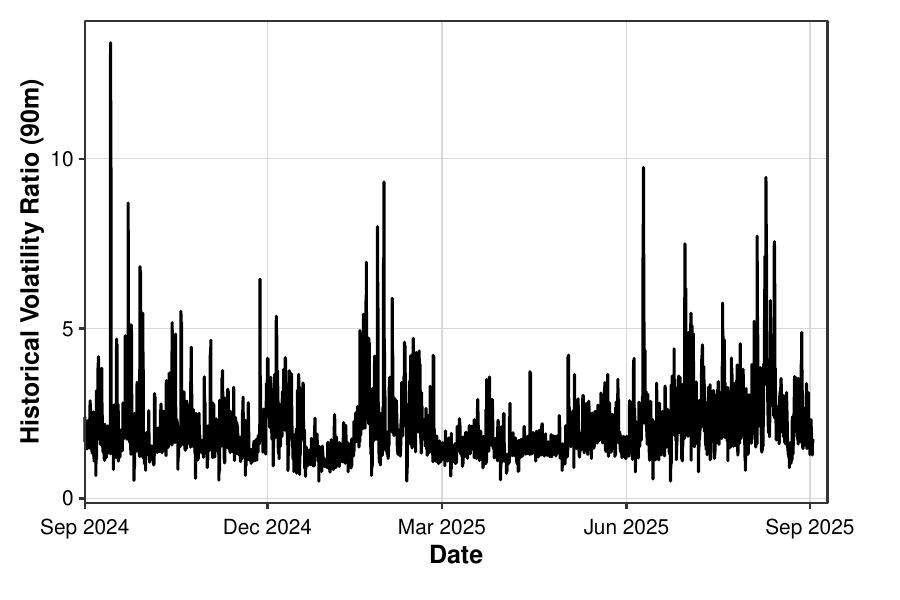} \\
    \end{tabular}
    \caption{Historical Volatility Ratio (HVR) — $h$-minute horizons (5, 10, 30, 90 minutes).}
    \label{fig:hvr-minutes}
\end{figure}

\begin{figure}[H]
    \centering
    \begin{tabular}{cc}
        \includegraphics[width=0.45\textwidth]{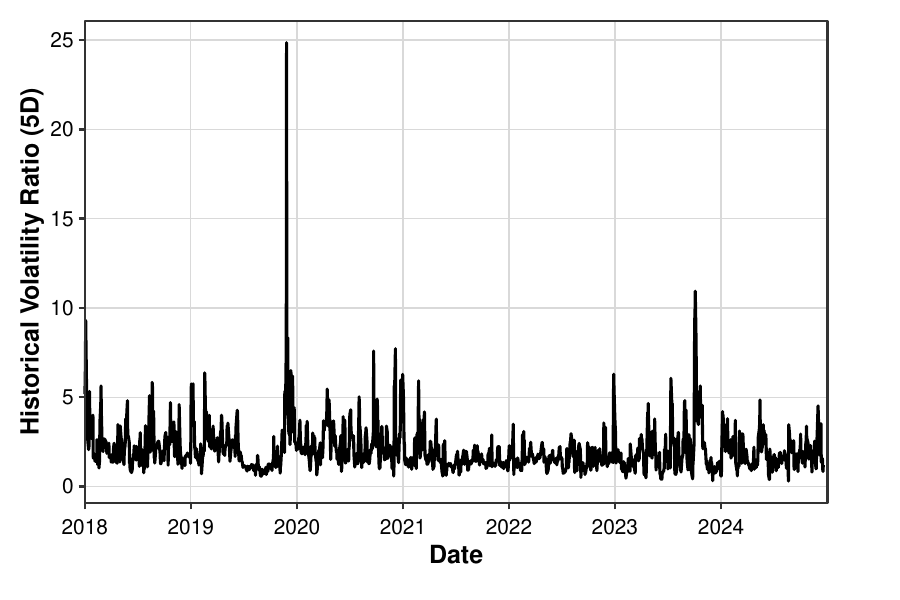}  &
        \includegraphics[width=0.45\textwidth]{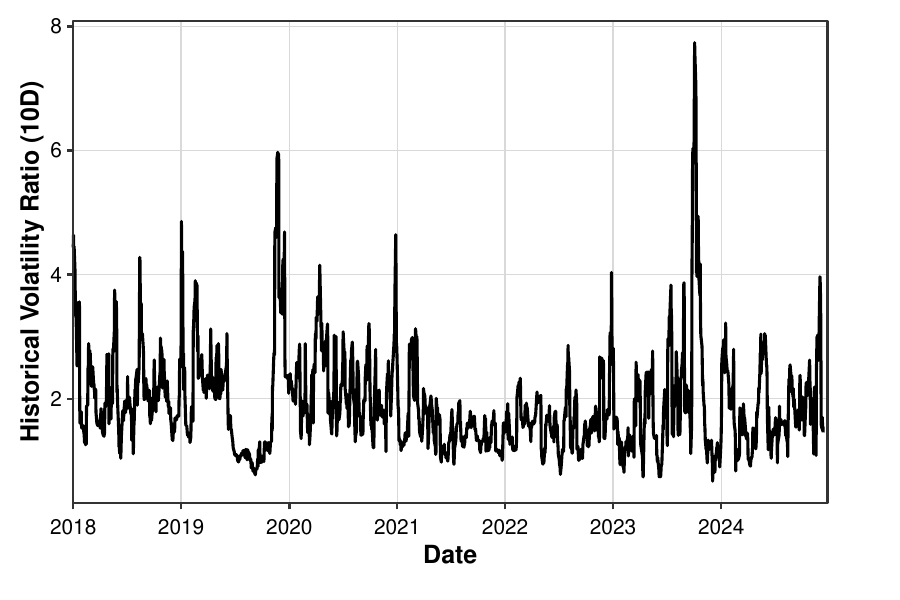} \\
        \includegraphics[width=0.45\textwidth]{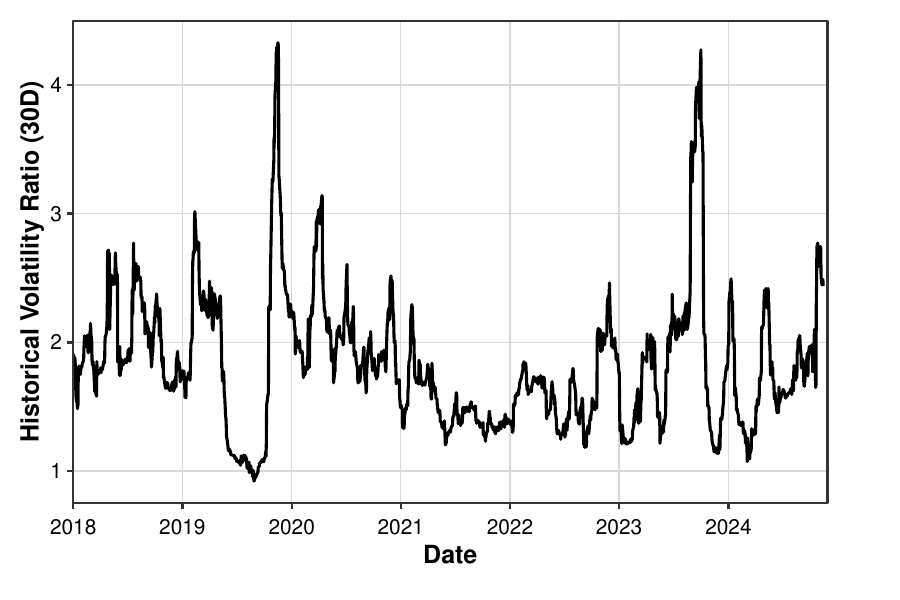} &
        \includegraphics[width=0.45\textwidth]{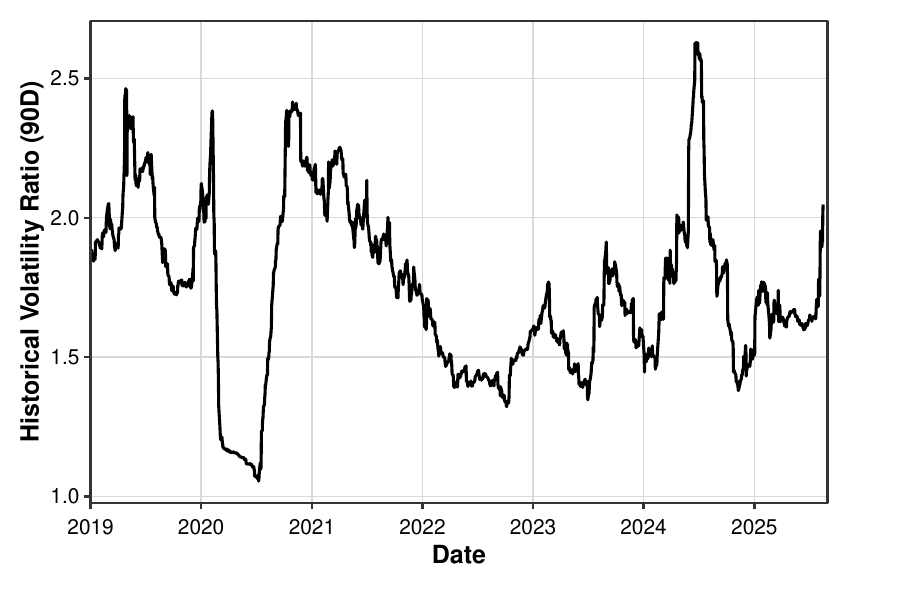} \\
    \end{tabular}
    \caption{Historical Volatility Ratio (HVR) — $h$-day horizons (5, 10, 30, 90 days).}
    \label{fig:hvr-daily}
\end{figure}

\hspace{1em}

From the graphs, we notice that the relative volatility of Apple Inc. does not show a trend, but rather temporary departures from the long-term equilibrium. This evidence supports the idea of using a measure of relative volatility (such as HVR and DVR) to better understand risk levels associated with the stock. Looking at these graphs, it also emerges that reversion to the equilibrium level appears faster for shorter horizons due to the mathematical nature of the estimation of rolling volatility (daily shocks tend to have a more persistent impact on wider rolling volatility windows).

\hspace{3em}

\subsection{Stationarity of HVR: ADF results}

We are now interested in testing whether these ratios show evidence of stationarity and whether the individual stock volatilities are cointegrated with the index. For this purpose, we performed an Augmented Dickey–Fuller (ADF) test on both the HVRs and the error terms of the Naive Volatility Model presented in Section 5.3 for all the stocks in the analysis across the examined horizons \citep{EngleGranger1987}. The share of series classified as stationary is \textbf{100.00\%} (5 minutes), \textbf{100.00\%} (10 minutes), \textbf{100.00\%} (30 minutes), \textbf{100.00\%} (90 minutes), \textbf{100.00\%} (5 days), \textbf{100.00\%} (10 days), \textbf{99.37\%} (30 days), and \textbf{22.38\%} (90 days). The share of series classified as cointegrated with the index volatility is \textbf{100.00\%} (5 minutes), \textbf{100.00\%} (10 minutes), \textbf{100.00\%} (30 minutes), \textbf{100.00\%} (90 minutes), \textbf{100.00\%} (5 days), \textbf{100.00\%} (10 days), \textbf{99.79\%} (30 days), and \textbf{84.10\%} (90 days).

\begin{table}[H]
\centering
\caption{ADF on HVR and Engle--Granger residuals (cointegration): percentage of series by decision}
\label{tab:vol_summary1}
\begin{threeparttable}
\begin{tabular}{lcc}
\toprule
\textbf{Horizon} & \textbf{Stationary (\%)} & \textbf{Cointegrated (\%)} \\
\midrule
5 minutes  & 100.00 & 100.00 \\
10 minutes & 100.00 & 100.00 \\
30 minutes & 100.00 & 100.00 \\
90 minutes & 100.00 & 100.00 \\
5 days  & 100.00 & 100.00 \\
10 days & 100.00 & 100.00 \\
30 days & 99.37  & 99.79 \\
90 days & 22.38  & 84.10 \\
\bottomrule
\end{tabular}
\begin{tablenotes}\footnotesize
\item Notes: ADF rejects the unit-root null at 5\% $\Rightarrow$ Stationary. Series with fewer than 25 observations were dropped.
\end{tablenotes}
\end{threeparttable}
\end{table}

\vspace{2em}

\justifying

The ADF test supports the hypothesis of mean reversion of the HVR and cointegration among volatilities for a sizable fraction of stocks at short horizons, but the evidence weakens over longer horizons. This pattern is consistent with the mechanics of rolling volatility: as the length of the rolling window increases, the estimator becomes a moving average of order $(n\!-\!1)$, so that the shocks dissipate more slowly and near-unit-root behavior becomes more plausible at long horizons.

\hspace{2em}

\subsection{HVR distribution}

In this section we analyze the distribution of the average HVR for each stock across all the $h$-minute and $h$-day horizons. Below we show the histograms of the distribution of average HVRs across the time horizons considered.

\begin{figure}[H]
    \centering
    \begin{tabular}{cc}
        \includegraphics[width=0.45\textwidth]{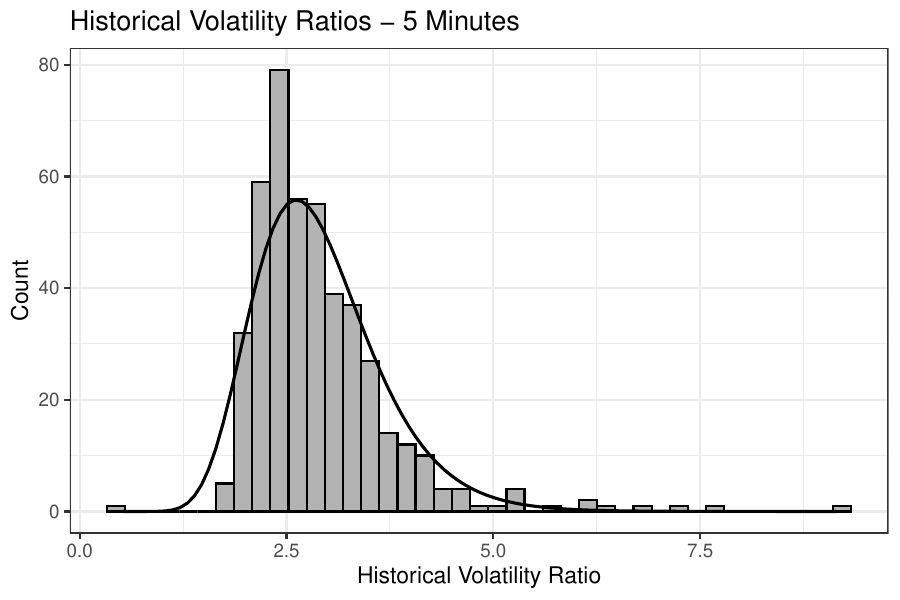}  &
        \includegraphics[width=0.45\textwidth]{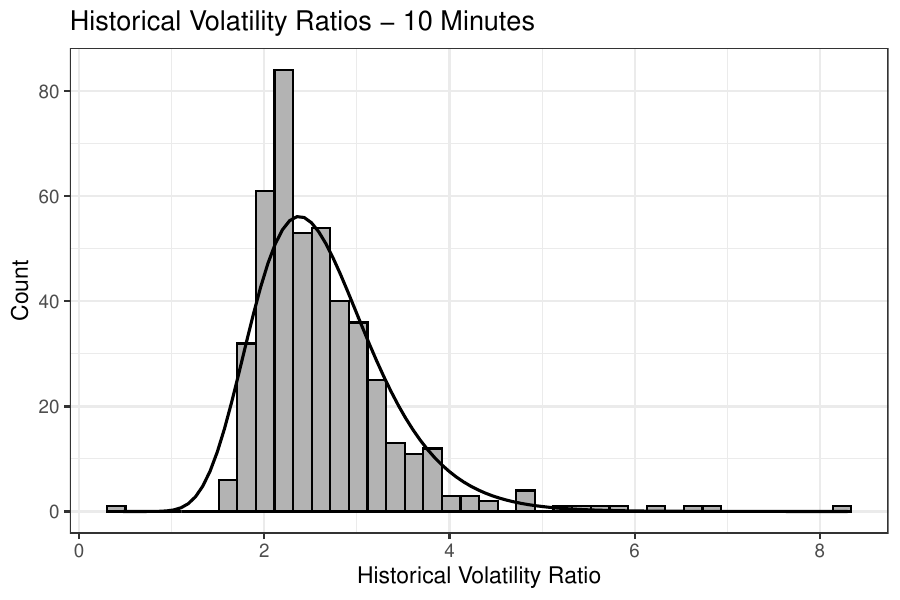} \\
        \includegraphics[width=0.45\textwidth]{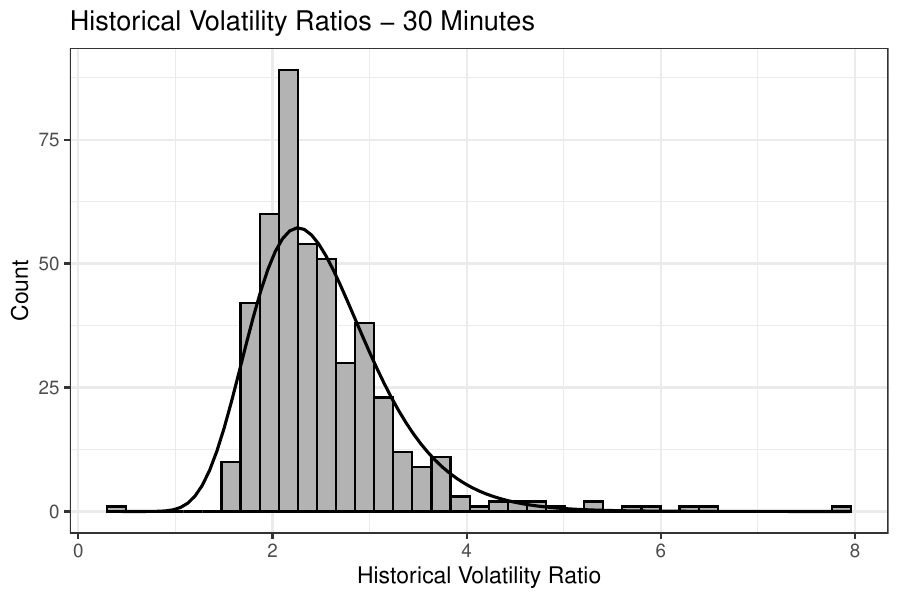} &
        \includegraphics[width=0.45\textwidth]{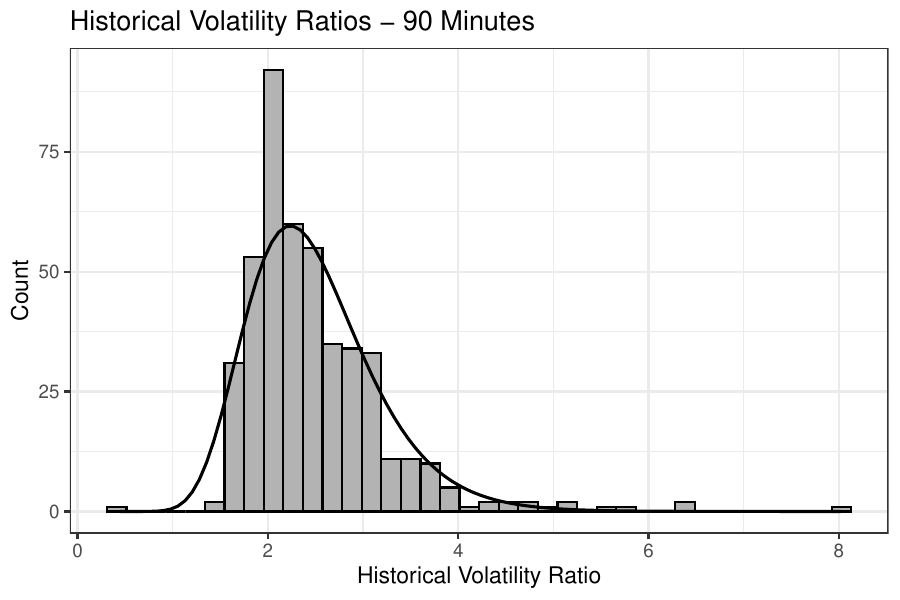} \\
    \end{tabular}
    \caption{Average Historical Volatility Ratio (HVR) — $h$-minute horizons.}
    \label{fig:hvr-dist-minute}
\end{figure}

\begin{figure}[H]
    \centering
    \begin{tabular}{cc}
        \includegraphics[width=0.45\textwidth]{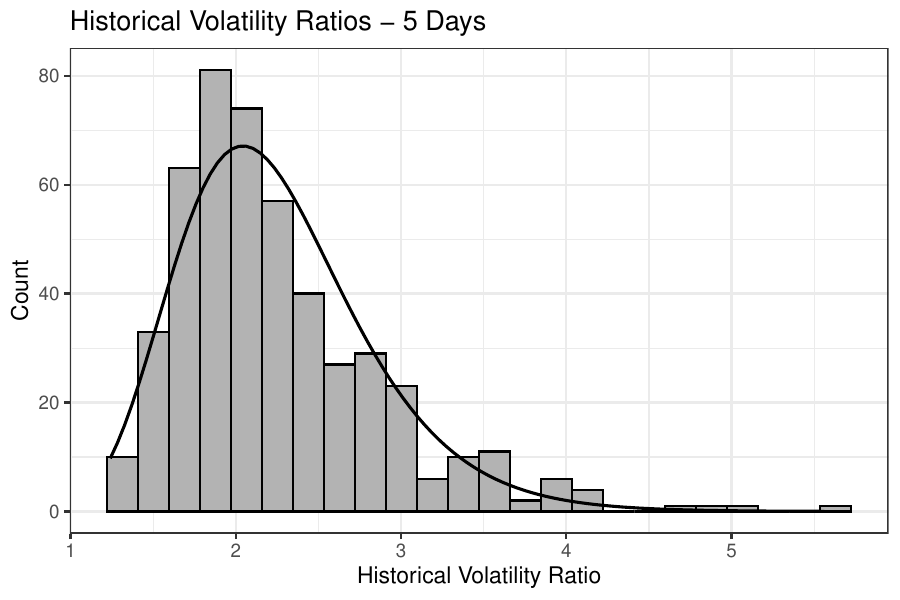}  &
        \includegraphics[width=0.45\textwidth]{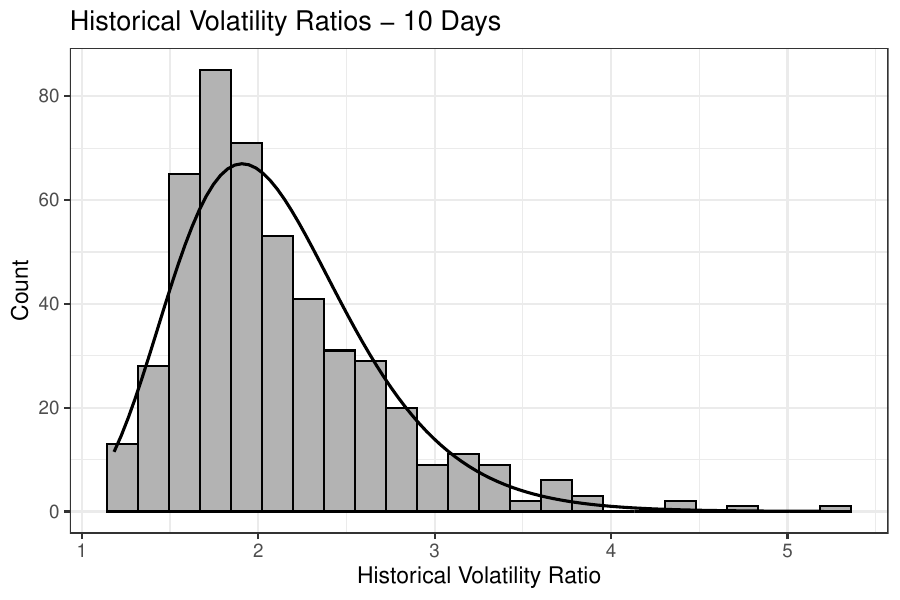} \\
        \includegraphics[width=0.45\textwidth]{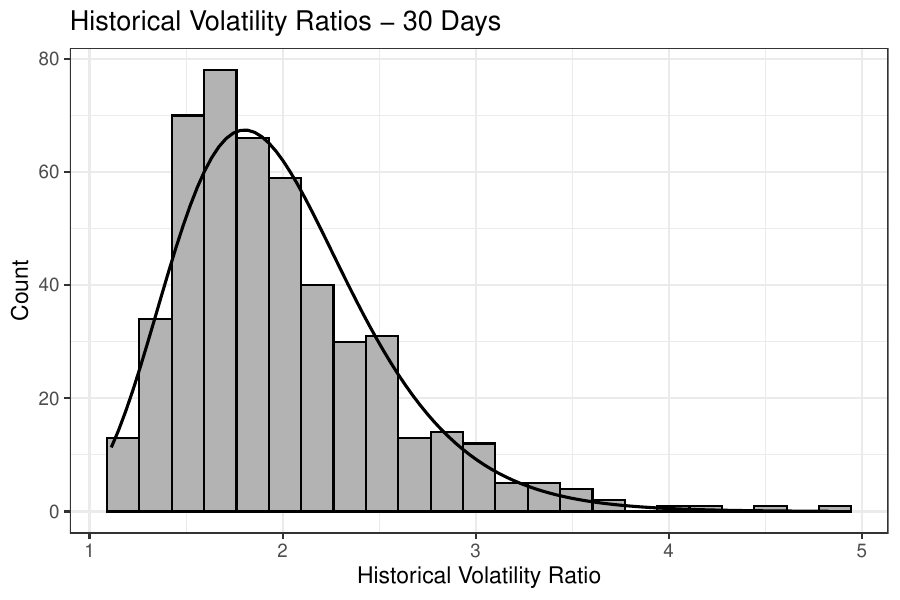} &
        \includegraphics[width=0.45\textwidth]{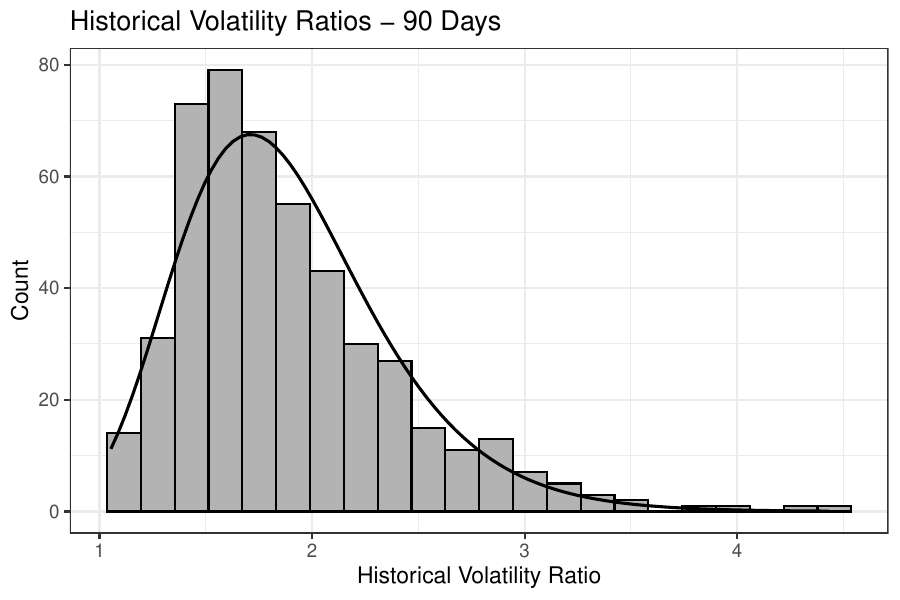} \\
    \end{tabular}
    \caption{Average Historical Volatility Ratio (HVR) — $h$-day horizons.}
    \label{fig:hvr-dist-daily}
\end{figure}

These distributions can be approximated by a log-normal curve as displayed in the graphs. As we test for significance of the log-normality hypothesis, we notice that, while the data exhibit kurtosis levels that are consistent with log-normality, they also exhibit excessive skewness. Therefore, we suggest using a log-t distribution to better capture the excessive skewness when modeling these distributions. Below we report the QQ-plots for average HVRs at 5 and 90 days and at 5 and 90 minutes. QQ-plots for the other horizons are available upon request and are consistent with the patterns shown.

\vspace{4em}

\begin{figure}[H]
  \centering
  \begin{minipage}{0.48\linewidth}
    \centering
    \includegraphics[width=\linewidth]{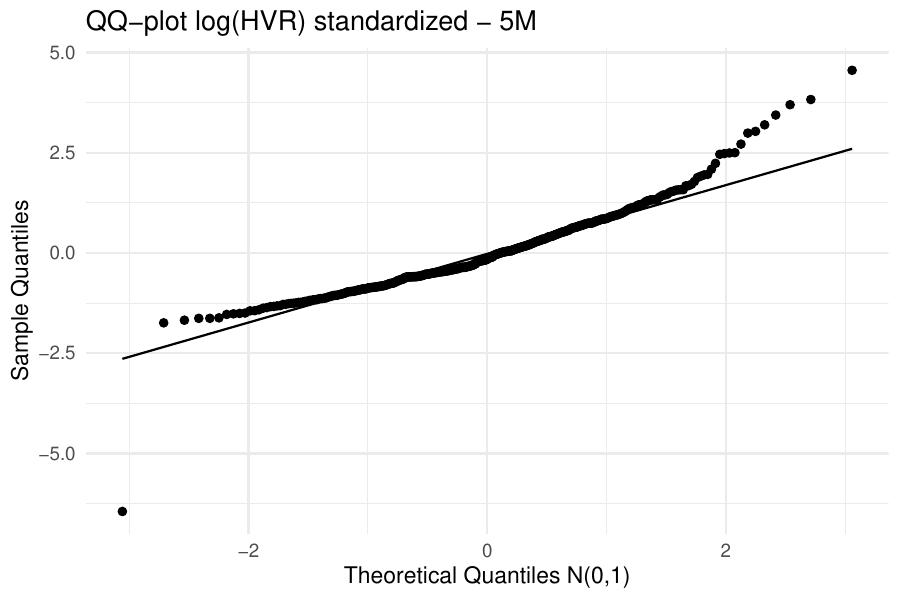}\\
    \small (a) 5 minutes
  \end{minipage}\hfill
  \begin{minipage}{0.48\linewidth}
    \centering
    \includegraphics[width=\linewidth]{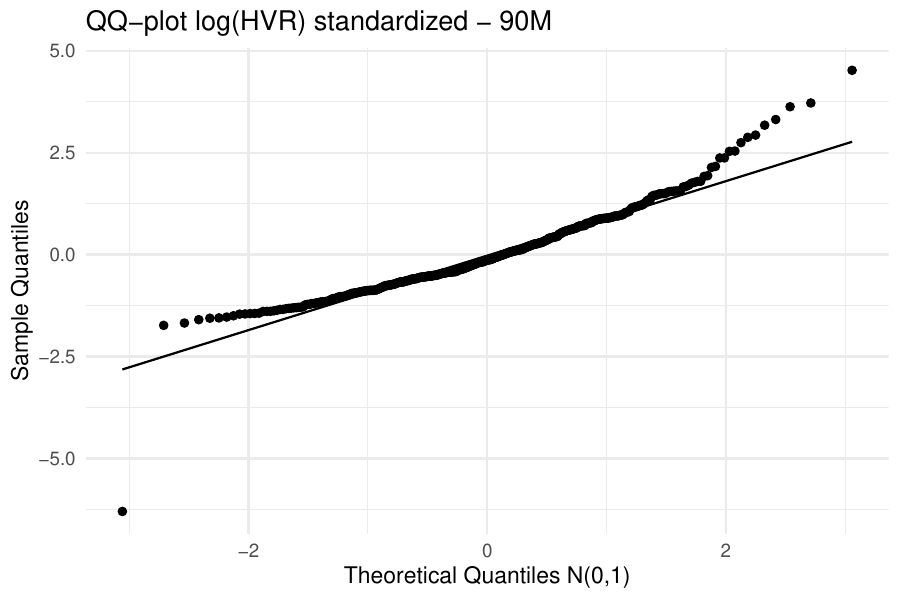}\\
    \small (b) 90 minutes
  \end{minipage}
  \caption{QQ-plots of standardized $\log(\mathrm{HVR})$ at 5 and 90 minutes.}
  \label{fig:qq-loghvr-5m-90m}
\end{figure}

\vspace{4em}

\begin{figure}[H]
  \centering
  \begin{minipage}{0.48\linewidth}
    \centering
    \includegraphics[width=\linewidth]{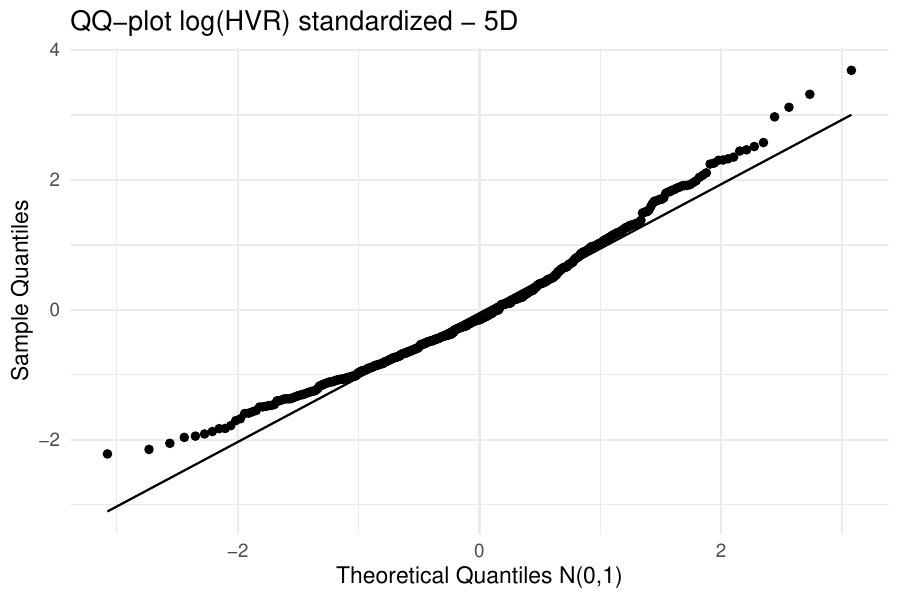}\\
    \small (a) 5 days
  \end{minipage}\hfill
  \begin{minipage}{0.48\linewidth}
    \centering
    \includegraphics[width=\linewidth]{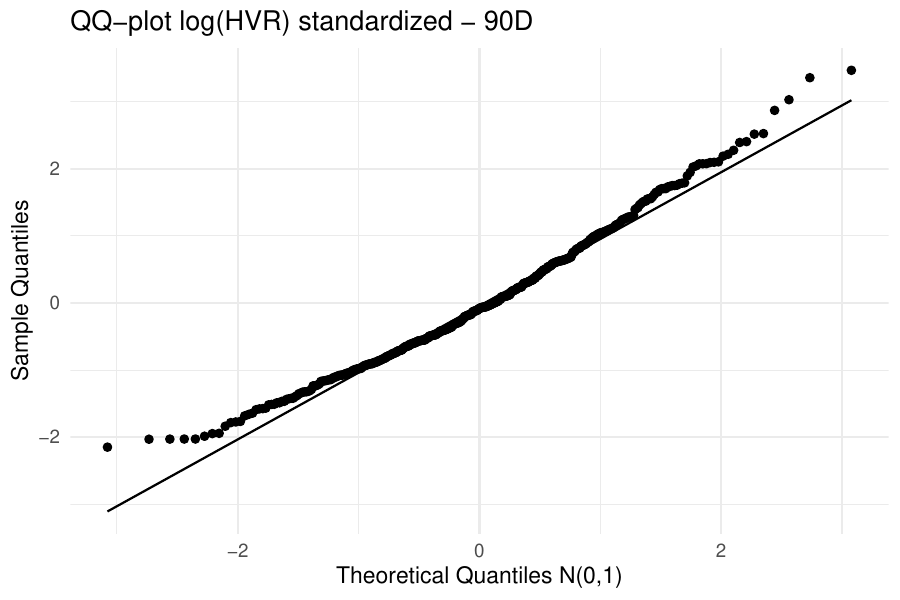}\\
    \small (b) 90 days
  \end{minipage}
  \caption{QQ-plots of standardized $\log(\mathrm{HVR})$ at 5 and 90 days.}
  \label{fig:qq-loghvr-5d-90d}
\end{figure}

\newpage

\subsection{ARIMA models for HVRs}

\vspace{1em}

We estimate univariate ARIMA models for each asset’s HVR at horizons $k\!\in\!\{5,10,30,90\}$ minutes and days using \texttt{auto.arima}. For each horizon, we (i) record the selected orders $(p,d,q)$ across all series, (ii) report the \emph{modal} value of each parameter ($p$, $d$, $q$) with its frequency share, (iii) identify the single most frequent \emph{triple} $(p,d,q)$ and its coverage, and (iv) re–estimate that triple uniformly on all series to obtain comparable in–sample errors (MAPE, RMSE). Results are in Table~\ref{tab:arima-summary}.

\vspace{2em}

\begin{table}[H]
\centering
\small
\setlength{\tabcolsep}{3.5pt}
\caption{ARIMA on HVR by horizon: modal parameter values, best triple coverage, and average in–sample errors}
\label{tab:arima-summary}
\begin{threeparttable}
\begin{tabularx}{\linewidth}{l c c c c c S[table-format=2.2] S[table-format=1.3]}
\toprule
\textbf{Horizon} &
\textbf{Modal $p$ (\%)} &
\textbf{Modal $d$ (\%)} &
\textbf{Modal $q$ (\%)} &
\textbf{Best $(p,d,q)$} &
\textbf{Coverage (\%)} &
\textbf{MAPE (\%)} &
\textbf{RMSE} \\
\midrule
5 minutes  & $p{=}5$ (43.1) & $d{=}1$ (100) & $q{=}5$ (93.5) & (5,1,5) & 42.4  & \multicolumn{1}{c}{--} & \multicolumn{1}{c}{--} \\
10 minutes  & $p{=}5$ (39.7) & $d{=}1$ (100) & $q{=}0$ (43.1) & (5,1,0) & 28.6  & \multicolumn{1}{c}{--} & \multicolumn{1}{c}{--} \\
30 minutes  & $p{=}1$ (28.6) & $d{=}1$ (100) & $q{=}0$ (51.8) & (1,1,0) & 15.4  & 3.97 & 0.294 \\
90 minutes  & $p{=}0$ (50.4) & $d{=}1$ (100) & $q{=}0$ (30.1) & (0,1,2) & 12.5  & 1.43 & 0.171 \\
5 days  & $p{=}1$ (23.4) & $d{=}1$ (50.8) & $q{=}5$ (29.7) & (5,0,2) & 7.5  & 33.60 & 1.150 \\
10 days & $p{=}0$ (27.2) & $d{=}1$ (60.9) & $q{=}0$ (33.3) & (0,1,0) & 19.5 & 11.80 & 0.502 \\
30 days & $p{=}0$ (41.2) & $d{=}1$ (81.6) & $q{=}0$ (32.4) & (0,1,0) & 27.4 &  3.63 & 0.163 \\
90 days & $p{=}1$ (52.5) & $d{=}1$ (96.2) & $q{=}1$ (47.9) & (1,1,1) & 36.2 &  1.23 & 0.056 \\
\bottomrule
\end{tabularx}

\begin{tablenotes}\footnotesize
\item ``Modal $p$ (\%)'' etc.\ report the single most frequent value of each parameter and its share among series at that horizon; these shares do not sum to 100 across values.
\item ``Best $(p,d,q)$'' is the most frequent triple from \texttt{auto.arima}; ``Coverage'' is its cross–sectional percentage.
\item Errors (MAPE, RMSE) are in–sample averages after re–estimating the best triple on each series at the given horizon.
\item Missing MAPE (\%) and RMSE for the 5- and 10-minute horizons are due to the presence of zeros in the series, which inflates these values toward infinity.
\end{tablenotes}
\end{threeparttable}
\end{table}

\vspace{2em}

From the table presented, it emerges a clear horizon pattern. At very short intraday horizons (5–10 minutes), the optimal ARIMA models show a high order of AR and MA terms (e.g., (5,1,5)/(5,1,0)), typical of microstructure-generated short-run dynamics. As we move to 30–90 minutes, the specification gets simpler ((1,1,0)/(0,1,2)) and errors are lower. Switching to daily frequency, we notice the same pattern as the parameters of the model become progressively more parsimonious. This behavior is again explained by the mathematical construction of rolling-window estimations. Overall, first-order differencing is a shared characteristic across all horizons and frequencies, but the high order of p and q terms becomes progressively unnecessary.

\vspace{2em}

\subsection{Performance: VECM vs. Classical Markowitz}

This section reports a summary of the performance of the Vector Error Correction Model (VECM) across the examined horizons and frequencies. The model is compared to the classical Markowitz covariance matrix, and the performance is evaluated using the MAPE. The test is performed by randomly selecting positive weights $(w \in (0,1))$ and assets for 100 portfolios and building both the VECM and the covariance matrix on the previous 1000 observations of returns. These models are then used to predict the volatility of each portfolio in the next $h$=\{5, 10, 30, 90\} observations, and the results are compared with the realized volatility of the portfolio in the same following $h$=\{5, 10, 30, 90\} observations. Below, we report the obtained results.

\begin{table}[H]
\centering
\small
\caption{\textbf{VECM vs Classical by horizon and portfolio size} (mean across 100 random portfolios per size)}
\label{tab:vecm-by-n}
\setlength{\tabcolsep}{3.5pt}
\makebox[\textwidth][c]{%
\begin{tabular}{l *{4}{cccc}}
\toprule
 & \multicolumn{4}{c}{\textbf{$N=10$}}
 & \multicolumn{4}{c}{\textbf{$N=30$}}
 & \multicolumn{4}{c}{\textbf{$N=50$}}
 & \multicolumn{4}{c}{\textbf{$N=80$}} \\
\cmidrule(lr){2-5}\cmidrule(lr){6-9}\cmidrule(lr){10-13}\cmidrule(lr){14-17}
\textbf{Horizon} & \textbf{V} & \textbf{C} & \(\boldsymbol{\Delta}\) & \textbf{Win\%}
                 & \textbf{V} & \textbf{C} & \(\boldsymbol{\Delta}\) & \textbf{Win\%}
                 & \textbf{V} & \textbf{C} & \(\boldsymbol{\Delta}\) & \textbf{Win\%}
                 & \textbf{V} & \textbf{C} & \(\boldsymbol{\Delta}\) & \textbf{Win\%} \\
\midrule
5 minutes   & 48.54 & 117.82 & 69.28 & 70.0 & 45.22 & 110.39 & 65.17 & 69.0 & 46.73 & 122.80 & 76.07 & 77.0 & 46.59 & 125.24 & 78.65 & 80.0 \\
10 minutes  & 38.36 & 98.63 & 60.27 & 78.0 & 34.39 & 91.94 & 57.55 & 77.0 & 46.05 & 101.11 & 55.06 & 70.0 & 37.35 & 89.22 & 51.87 & 70.0 \\
30 minutes  & 37.09 & 72.66 & 35.57 & 66.0 & 47.65 & 73.44 & 25.79 & 67.0 & 40.83 & 88.42 & 47.59 & 75.0 & 50.67 & 78.98 & 28.31 & 65.0 \\
90 minutes  & 55.22 & 62.09 & 6.87 & 51.0 & 70.65 & 71.47 & 0.82 & 48.0 & 72.68 & 74.02 & 1.34 & 47.0 & 70.48 & 62.28 & -8.20 & 39.0 \\
5 days      & 46.98 & 93.83 & 46.85 & 71.0 & 43.47 & 94.59 & 51.12 & 80.0 & 50.20 & 98.83 & 48.63 & 77.0 & 58.15 & 105.75 & 47.60 & 72.0 \\
10 days     & 28.04 & 58.65 & 30.61 & 72.0 & 38.31 & 72.70 & 34.39 & 74.0 & 35.93 & 76.25 & 40.32 & 80.0 & 44.93 & 77.45 & 32.52 & 77.0 \\
30 days     & 30.68 & 55.92 & 25.24 & 75.0 & 34.52 & 66.62 & 32.10 & 80.0 & 37.62 & 64.51 & 26.89 & 78.0 & 41.76 & 64.96 & 23.20 & 70.0 \\
90 days     & 29.79 & 53.14 & 23.35 & 77.0 & 36.17 & 61.80 & 25.63 & 77.0 & 42.24 & 69.21 & 26.97 & 79.0 & 41.68 & 61.01 & 19.33 & 68.0 \\
\bottomrule
\end{tabular}%
}
\begin{flushleft}\footnotesize
Notes: V = mean MAPE (VECM); C = mean MAPE (Classical); \(\Delta=\)C\(-\)V in percentage points (positive favors VECM); Win\% = share of portfolios with MAPE\(_\text{VECM}<\)MAPE\(_\text{Classical}\).
\end{flushleft}
\end{table}

From this table, it emerges a clear superiority of the VECM over the classical covariance matrix estimation. At the same time, we notice that the difference in accuracy between these two models tends to dissipate at longer horizons both for minute and daily data; this decay is faster for the minute data. This result is not surprising, as the cointegration and stationarity hypotheses are more fragile on longer horizons. The model, therefore, shows a greater ability to capture existing volatility structures for shorter frequencies and tends to become unnecessarily complex after a certain horizon.

When computing the VECM with the assumption of cointegration and with variables close to zero, the result tends to be unstable for certain horizons. For this reason, when fitting the model, we impose a "guardrail" that sets the expected volatility of the portfolio equal to its average realized volatility when the VECM yields an expectation greater than three times the average volatility of the portfolio in the previous $h$ periods (minutes for intraday, days for daily). The table below reports the number of times that the "guardrail" was triggered. We notice greater instability of the model for longer forecast horizons at minute and daily frequencies, consistent with the results obtained in Table~\ref{tab:vecm-by-n}.

\begin{table}[H]
\centering
\small
\setlength{\tabcolsep}{3.5pt}
\caption{VECM number of guardrail hits on 100 simulations}
\label{tab:guardrail-hits}
\begin{threeparttable}
\begin{tabularx}{\linewidth}{l *{4}{>{\centering\arraybackslash}X}}
\toprule
\textbf{Horizon} & \textbf{$N=10$} & \textbf{$N=30$} & \textbf{$N=50$} & \textbf{$N=80$} \\
\midrule
5 minutes  & 3.00 & 2.00 & 2.00 & 2.00 \\
10 minutes & 0.00 & 0.00 & 0.00 & 1.00 \\
30 minutes & 0.00 & 1.00 & 1.00 & 1.00 \\
90 minutes & 1.00 & 4.00 & 6.00 & 11.00 \\
5 days     & 0.00 & 1.00 & 3.00 & 2.00 \\
10 days    & 0.00 & 0.00 & 1.00 & 0.00 \\
30 days    & 0.00 & 0.00 & 0.00 & 4.00 \\
90 days    & 0.00 & 0.00 & 1.00 & 4.00 \\
\bottomrule
\end{tabularx}

\begin{tablenotes}\footnotesize
\item This table presents the number of "guardrail hits" for each horizon and $N$-stock portfolios.
\end{tablenotes}
\end{threeparttable}
\end{table}

\medskip
\noindent\textit{Boxplot evidence.} To benchmark forecasting accuracy across portfolio sizes, we report boxplots of the Absolute Percentage Error (APE) for VECM and the classical covariance model at horizons 5M, 10M, 30M, and 90M. For daily data: 5D, 10D, 30D, and 90D, using 100 random portfolios for each size $N\in\{10,30,50,80\}$. The boxes display the median and interquartile range, whiskers extend to $1.5\times$IQR, and outliers are shown as points.

\begin{figure}[H]
  \centering
  \begin{minipage}{0.49\linewidth}
    \centering
    \includegraphics[width=\linewidth,height=0.45\linewidth]{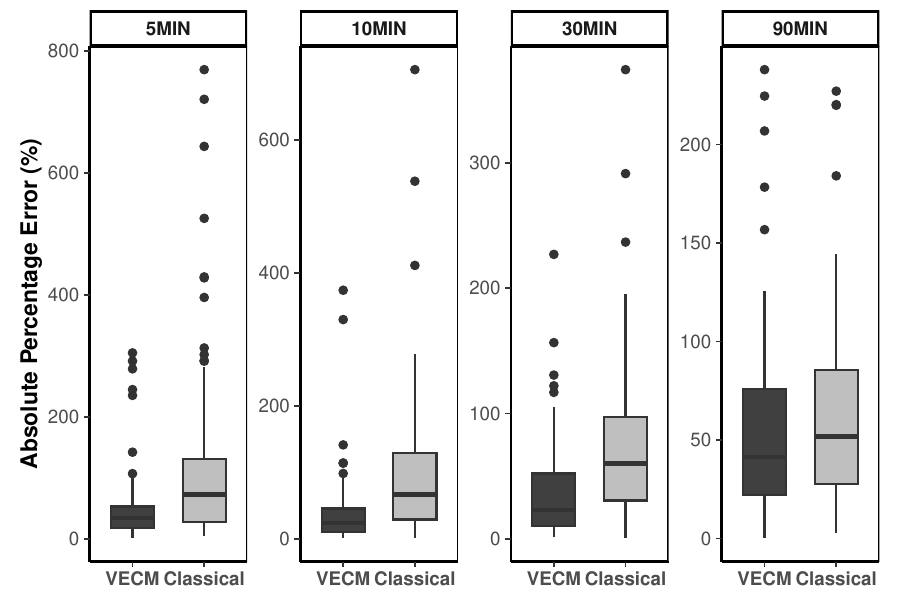}\\
    \small (a) $N=10$
  \end{minipage}\hfill
  \begin{minipage}{0.49\linewidth}
    \centering
    \includegraphics[width=\linewidth,height=0.45\linewidth]{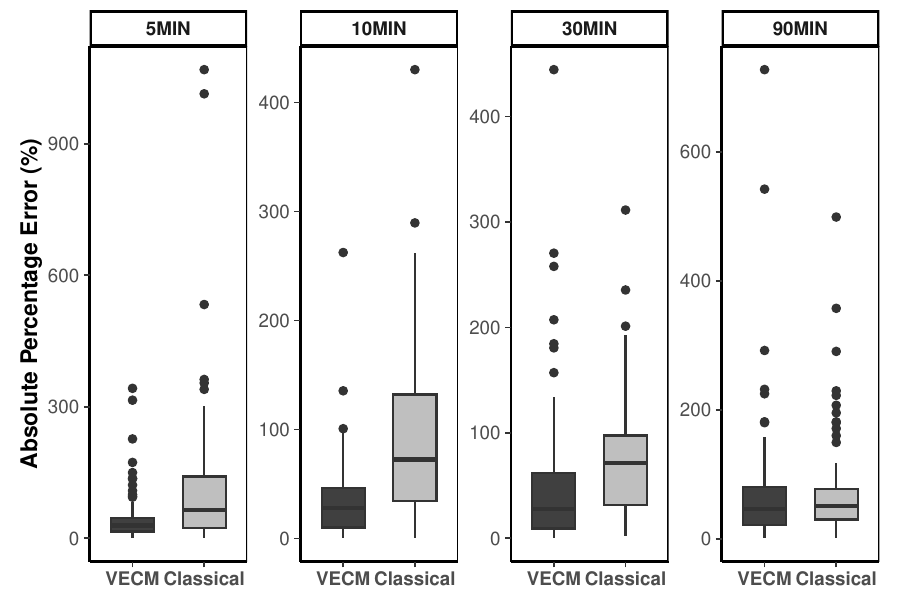}\\
    \small (b) $N=30$
  \end{minipage}\\[0.7em]
  \begin{minipage}{0.49\linewidth}
    \centering
    \includegraphics[width=\linewidth,height=0.45\linewidth]{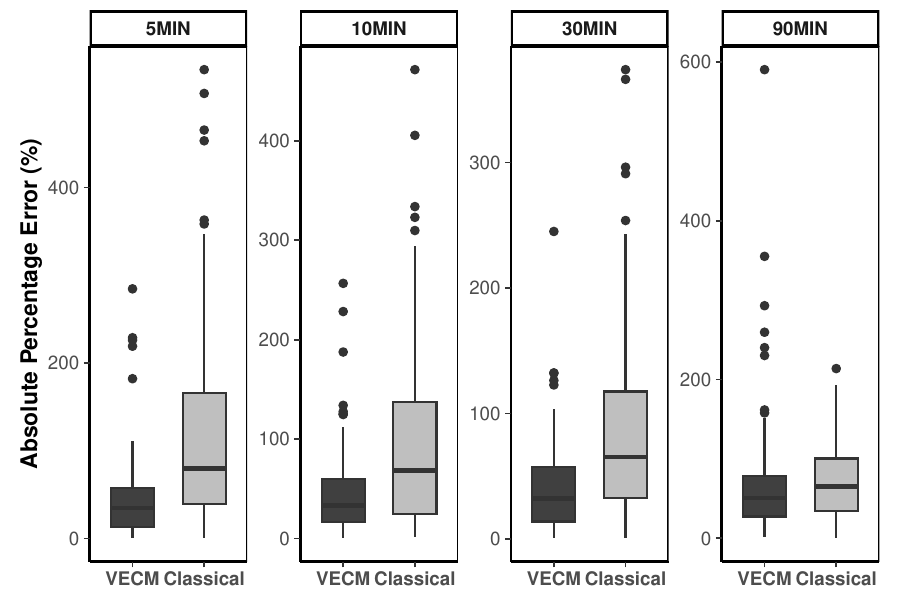}\\
    \small (c) $N=50$
  \end{minipage}\hfill
  \begin{minipage}{0.49\linewidth}
    \centering
    \includegraphics[width=\linewidth,height=0.45\linewidth]{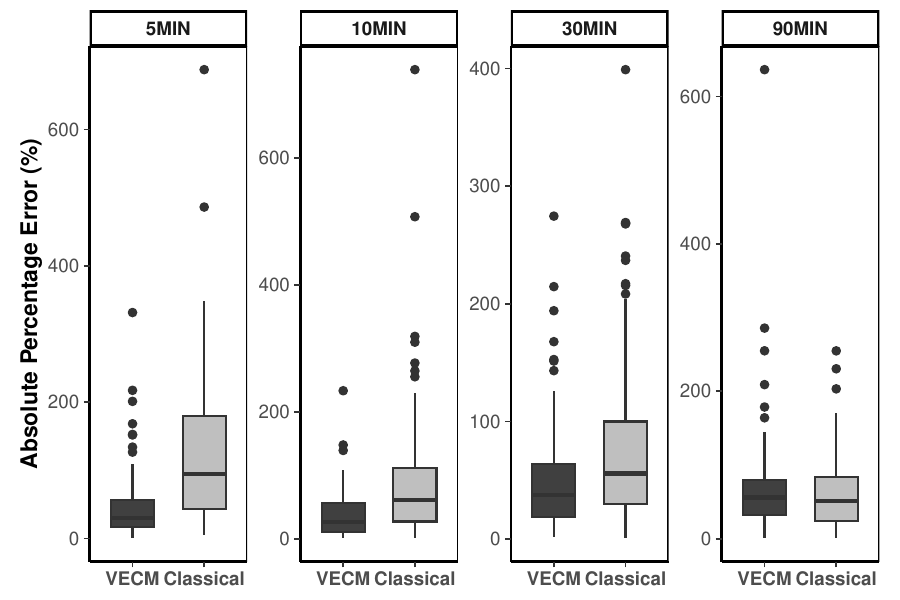}\\
    \small (d) $N=80$
  \end{minipage}
  \caption{APE boxplots by horizon (5M, 10M, 30M, 90M) for VECM vs.\ Classical across portfolio sizes.}
  \label{fig:ape-boxplots-by-n-min}
\end{figure}

\begin{figure}[H]
  \centering
  \begin{minipage}{0.49\linewidth}
    \centering
    \includegraphics[width=\linewidth,height=0.45\linewidth]{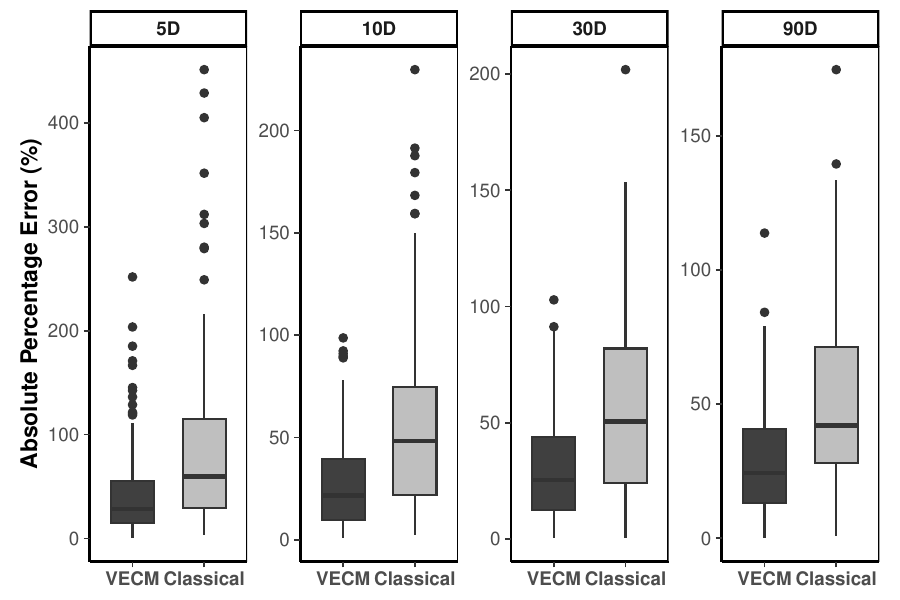}\\
    \small (a) $N=10$
  \end{minipage}\hfill
  \begin{minipage}{0.49\linewidth}
    \centering
    \includegraphics[width=\linewidth,height=0.45\linewidth]{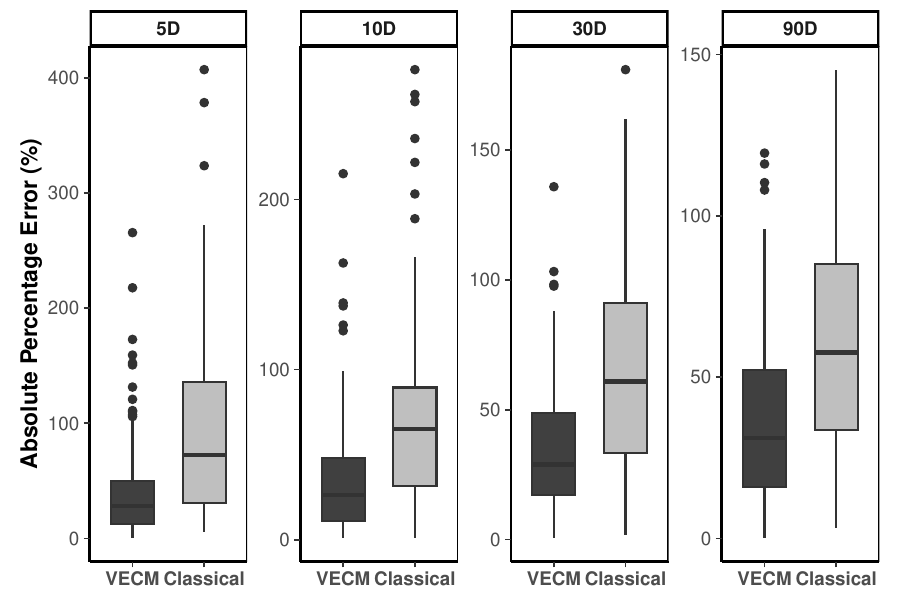}\\
    \small (b) $N=30$
  \end{minipage}\\[0.7em]
  \begin{minipage}{0.49\linewidth}
    \centering
    \includegraphics[width=\linewidth,height=0.45\linewidth]{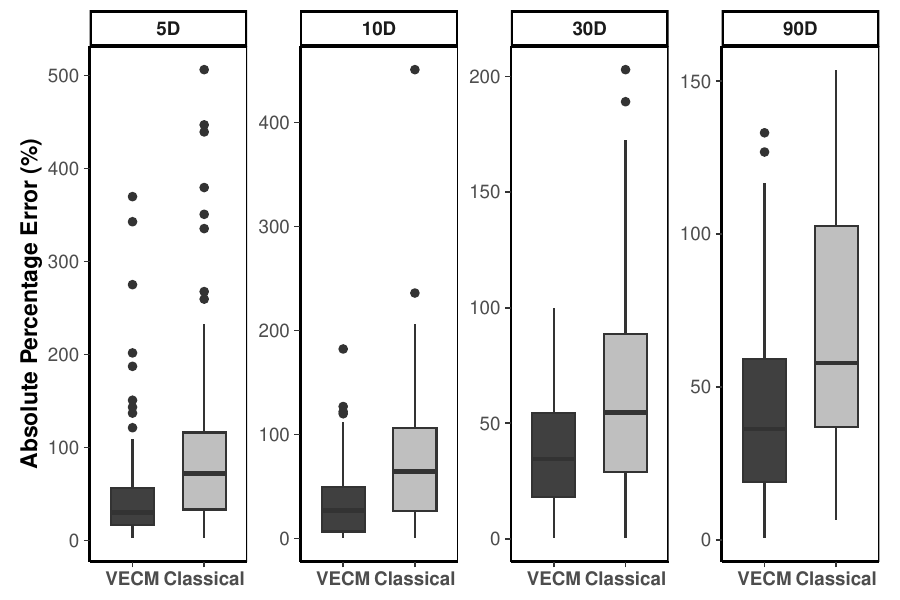}\\
    \small (c) $N=50$
  \end{minipage}\hfill
  \begin{minipage}{0.49\linewidth}
    \centering
    \includegraphics[width=\linewidth,height=0.45\linewidth]{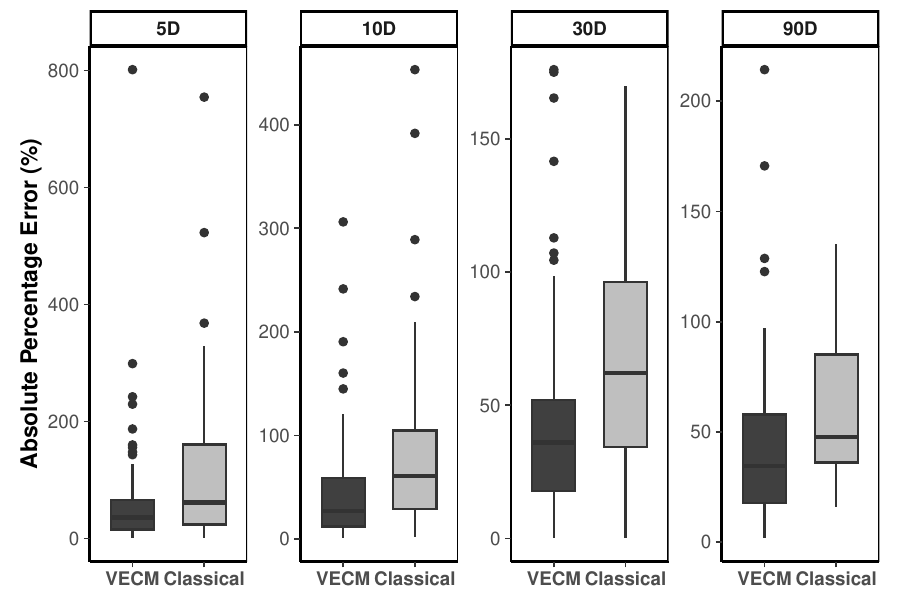}\\
    \small (d) $N=80$
  \end{minipage}
  \caption{APE boxplots by horizon (5D, 10D, 30D, 90D) for VECM vs.\ Classical across portfolio sizes.}
  \label{fig:ape-boxplots-by-n-day}
\end{figure}

\newpage

\section{Concluding Remarks}

\vspace{1.5em}

This paper has established that portfolio risk modeling with cointegration-based dynamics among volatilities has an empirical foundation and a clear operational advantage. Across S\&P 500 components and a wide range of horizons, we find widespread stationarity of HVR and strong cointegration between single-stock and benchmark volatilities.

\vspace{1em}

For forecasting, the VECM framework systematically outperforms the traditional covariance approach on MAPE at intraday (5–30 minutes) and short-to-medium daily horizons (5–30 days), delivering sizable accuracy gains with high win rates across portfolio sizes. The advantage narrows at the longest horizons, where rolling measures mechanically embed more persistence and the cointegration signal weakens; in these cases, VECM still remains competitive. Occasional unstable predictions are rare and predictable, rising mainly at the longest forecast windows. The “guardrail” results indicate that failures are not structural to the framework but concentrate in regimes with very low volatility or when differenced log-vol processes approach local boundaries.

\vspace{1em}

Overall, these findings validate the central claim: relative, cointegration-aware volatility modeling yields more stable and interpretable risk estimates than direct covariance estimation, precisely when forward accuracy matters most (short horizons and higher-dimensional portfolios). In practice, expressing covariances as market volatility scaled by stationary HVR (or forward-looking DVR) and a separate correlation matrix decouples market level, relative exposure, and cross-sectional dependence, producing more accurate forecasts and simpler stress tests.

\vspace{1em}

Two immediate implications follow. First, portfolio construction and risk budgeting can be reframed around stationary relative-risk processes (HVR/DVR) and a slower-moving correlation structure, with VECM linking short-run adjustments to long-run equilibria. Second, parsimony is a virtue: as horizons lengthen, simpler ARIMA specifications for HVR and fewer VECM lags suffice, while at very short horizons richer short-run dynamics can be accommodated without abandoning the error-correction discipline.

Limitations remain, such as the horizon-specific fragility of cointegration and the residual dependence on correlation estimates in high dimension, but the evidence suggests these are tractable with standard controls (e.g., correlation shrinkage, horizon-specific rank selection, and stability guardrails). Future work should extend the framework across asset classes and markets and develop an integrated framework that can capture the short-term volatility structures and equilibria across assets.

\newpage
\bibliographystyle{apalike}
\bibliography{references}

\begin{thebibliography}{}

\bibitem[Andersen et~al., 2001]{Andersen2001}
Andersen, T.~G., Bollerslev, T., Diebold, F.~X., and Labys, P. (2001).
\newblock The distribution of realized exchange rate volatility.
\newblock {\em Journal of the American Statistical Association}, 96(453):42--55.

\bibitem[Bollerslev, 1986]{Bollerslev1986}
Bollerslev, T. (1986).
\newblock Generalized autoregressive conditional heteroskedasticity.
\newblock {\em Journal of Econometrics}, 31(3):307--327.

\bibitem[Engle, 2002]{Engle2002}
Engle, R.~F. (2002).
\newblock Dynamic conditional correlation: A simple class of multivariate {GARCH} models.
\newblock {\em Journal of Business \& Economic Statistics}, 20(3):339--350.

\bibitem[Engle and Granger, 1987]{EngleGranger1987}
Engle, R.~F. and Granger, C. W.~J. (1987).
\newblock Co-integration and error correction: Representation, estimation, and testing.
\newblock {\em Econometrica}, 55(2):251--276.

\bibitem[Johansen, 1991]{Johansen1991}
Johansen, S. (1991).
\newblock Estimation and hypothesis testing of cointegration vectors in gaussian vector autoregressive models.
\newblock {\em Econometrica}, 59(6):1551--1580.

\bibitem[Ledoit and Wolf, 2003]{LedoitWolf2003}
Ledoit, O. and Wolf, M. (2003).
\newblock Improved estimation of the covariance matrix of stock returns with an application to portfolio selection.
\newblock {\em Journal of Empirical Finance}, 10(5):603--621.

\bibitem[Ledoit and Wolf, 2004]{LedoitWolf2004}
Ledoit, O. and Wolf, M. (2004).
\newblock A well-conditioned estimator for large-dimensional covariance matrices.
\newblock {\em Journal of Multivariate Analysis}, 88(2):365--411.

\bibitem[Markowitz, 1952]{Markowitz1952}
Markowitz, H. (1952).
\newblock Portfolio selection.
\newblock {\em The Journal of Finance}, 7(1):77--91.

\bibitem[Michaud, 1989]{Michaud1989}
Michaud, R.~O. (1989).
\newblock The markowitz optimization enigma: Is ``optimized'' optimal?
\newblock {\em Financial Analysts Journal}, 45(1):31--42.

\end{thebibliography}
\end{document}